\documentstyle[twoside,fleqn,espcrc2,epsfig]{article}
\title{Quantum chaos and QCD at finite chemical potential}
\author{H.~Markum\address{Institut f\"ur Kernphysik, Technische
  Universit\"at Wien, A-1040 Wien, Austria}, 
  R.~Pullirsch$^{\rm a}$\thanks{Poster presented by R.~Pullirsch}, 
  K.~Rabitsch$^{\rm a}$, and 
  T.~Wettig\address{Institut f\"ur Theoretische Physik, 
  Technische Universit\"at M\"unchen, D-85747 Garching, Germany}}
\begin{document}
\begin{abstract}
  We investigate the distribution of the spacings of adjacent
  eigenvalues of the lattice Dirac operator.  At zero chemical
  potential $\mu$, the nearest-neighbor spacing distribution $P(s)$
  follows the Wigner surmise of random matrix theory both in the
  confinement and in the deconfinement phase.  This is indicative of
  quantum chaos.  At nonzero chemical potential, the eigenvalues of
  the Dirac operator become complex.  We discuss how $P(s)$ can be
  defined in the complex plane.  Numerical results from an SU(3)
  simulation with staggered fermions are compared with predictions
  from non-hermitian random matrix theory, and agreement with the
  Ginibre ensemble is found for $\mu\approx 0.7$.
\end{abstract}
\date{\today}
\maketitle

\section{Introduction}
\label{sec1}
\vspace{-2mm}
\begin{figure*}
  \label{fig1}
  \hspace*{12mm}$\mu=0.1$\hspace*{20.3mm}$\mu=0.4$
  \hspace*{20.3mm}$\mu=0.7$\hspace*{20.3mm}$\mu=1.0$
  \hspace*{20.3mm}$\mu=2.2$\\[1mm]
  \centerline{\epsfig{figure=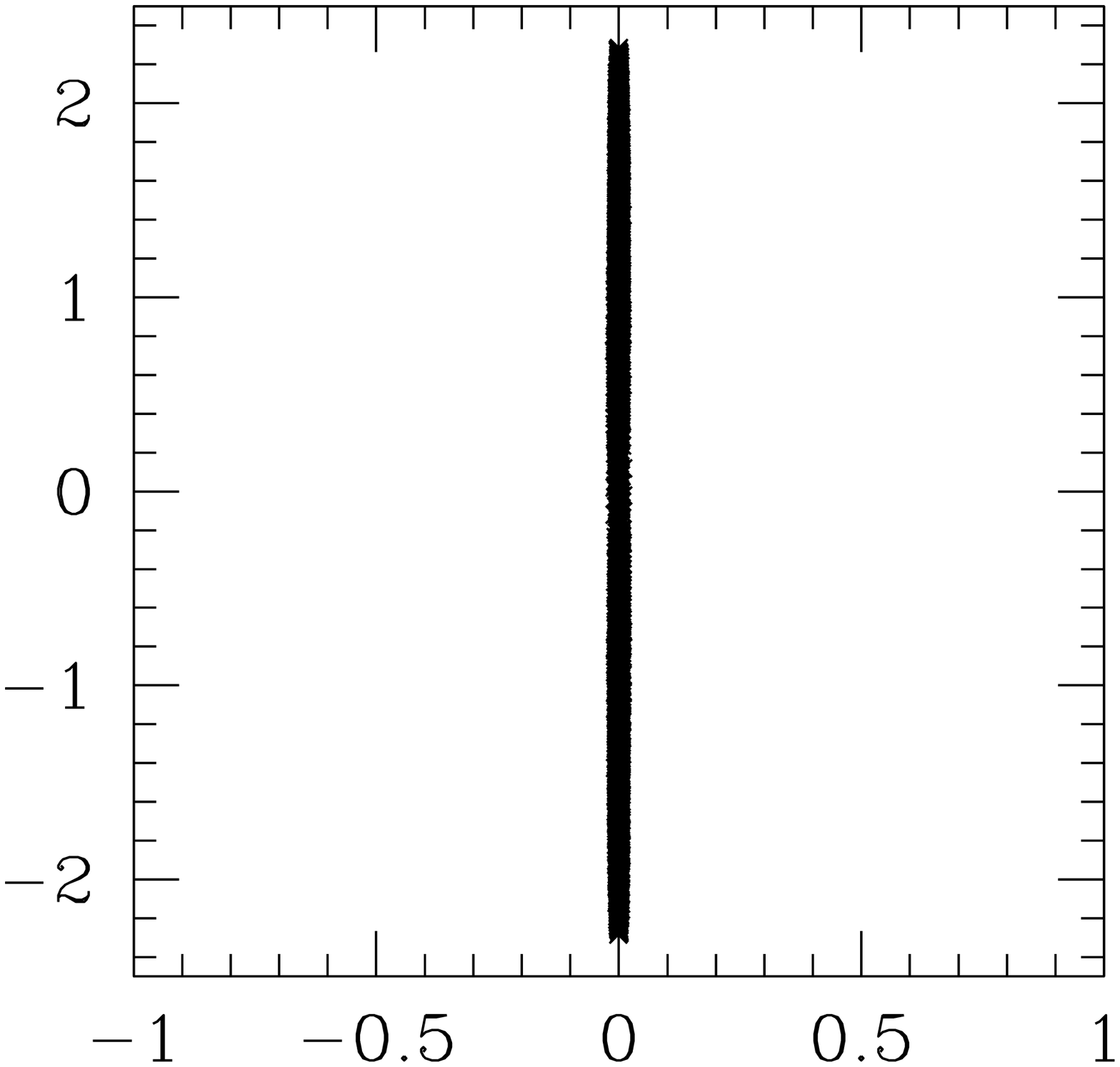,width=30mm}\hspace*{1mm}
    \epsfig{figure=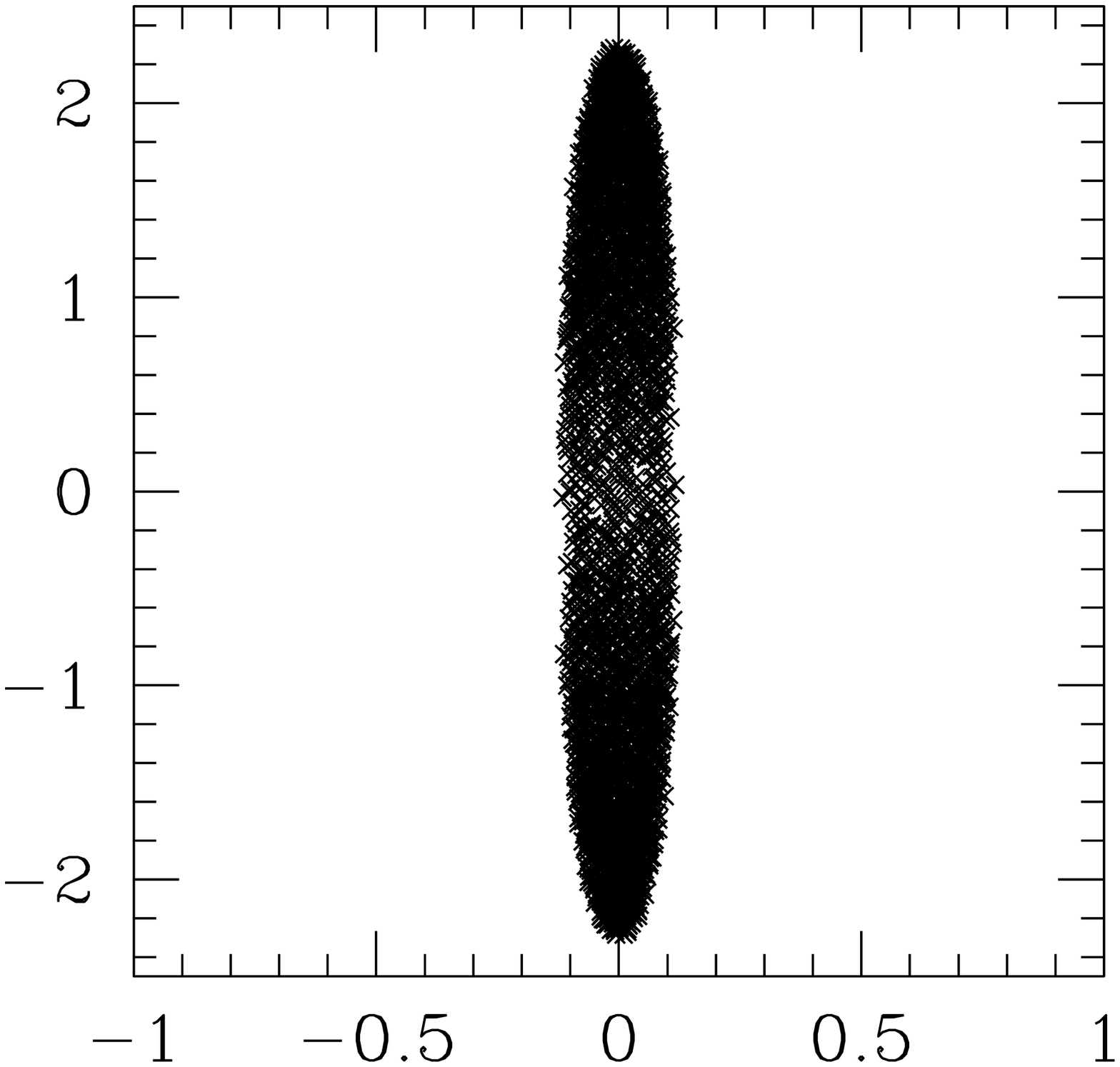,width=30mm}\hspace*{1mm}
    \epsfig{figure=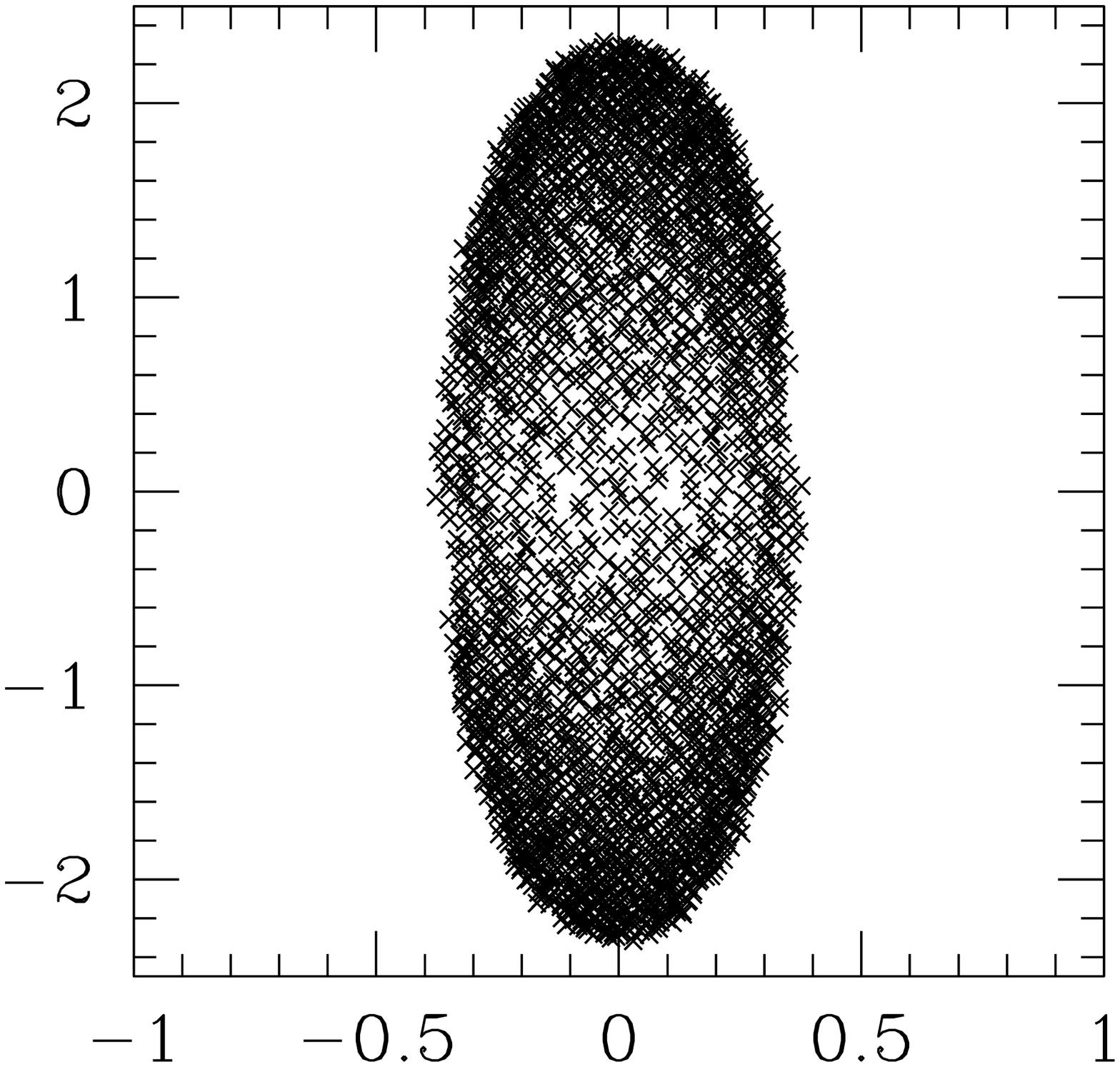,width=30mm}\hspace*{1mm}
    \epsfig{figure=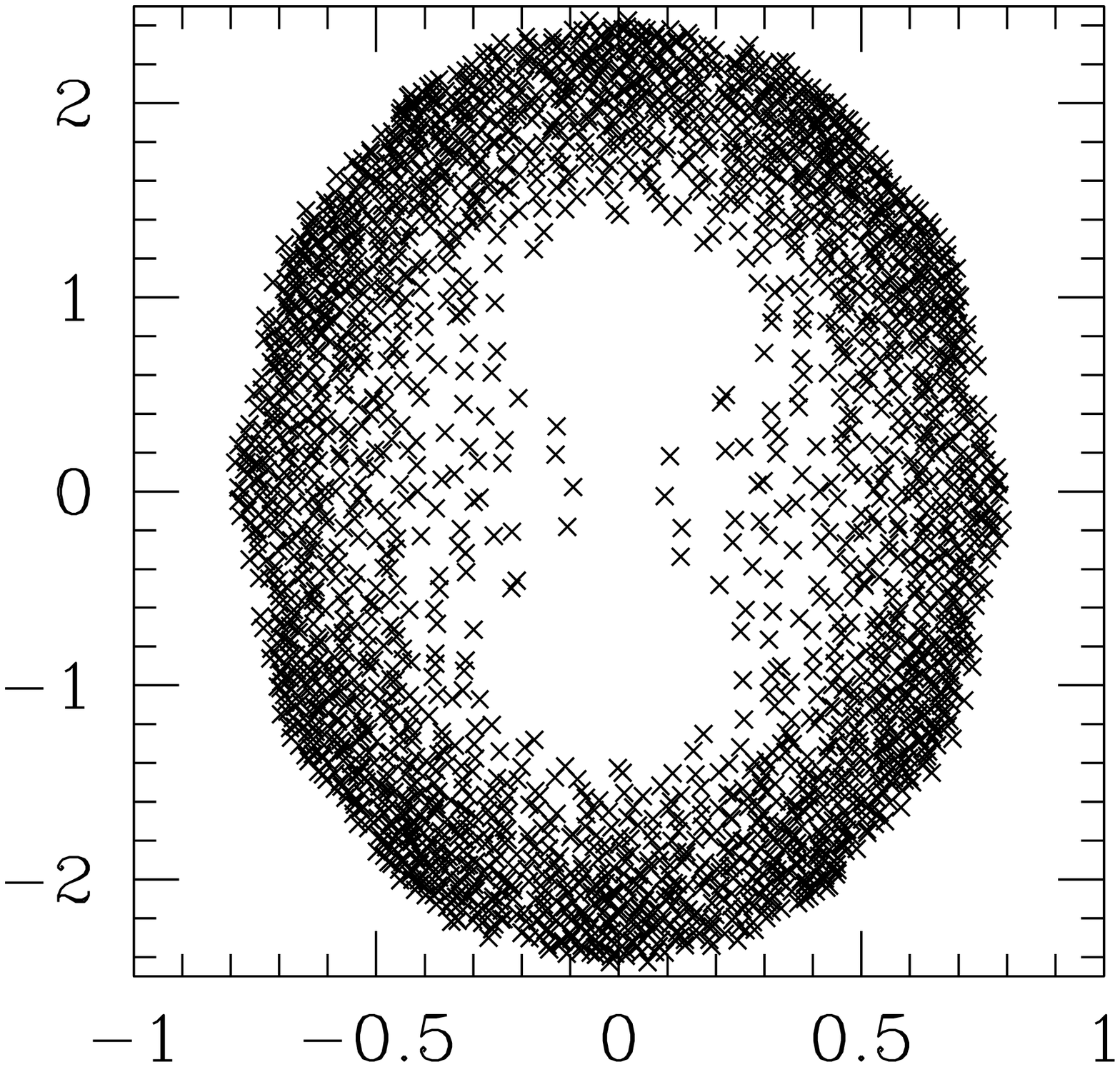,width=30mm}\hspace*{1mm}
    \epsfig{figure=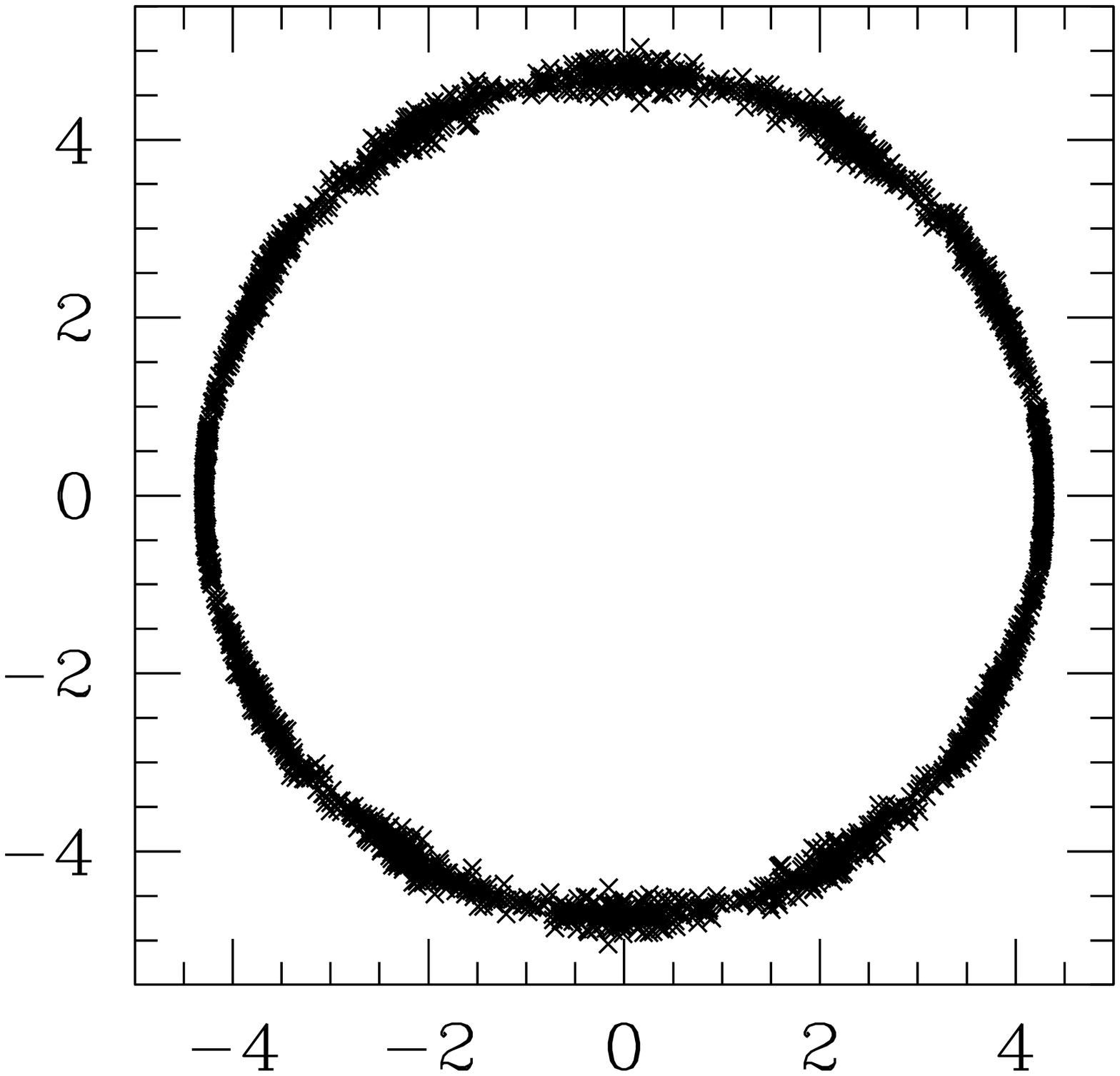,width=30mm}}
  \vspace*{-12mm}
  \caption{Complex eigenvalues of the Dirac operator at various values 
    of $\mu$ for a typical equilibrium configuration of QCD with staggered
    quarks without chemical potential (horizontal axes = real
    parts, vertical axes = imaginary parts, in units of $1/a$).}
\end{figure*}

The spectrum of the QCD Dirac operator, both in the continuum and on
the lattice, has several universal properties.  By ``universal'' one
means ``independent of the details of the dynamics'', e.g., independent
of the precise values of the simulation parameters on the lattice.
Such universal features can be described by random matrix theory
(RMT).  In this contribution, we are concerned with the eigenvalue
fluctuations in the bulk of the spectrum on the scale of the local
mean level spacing, measured by the distribution $P(s)$ of
the spacings $s$ of adjacent eigenvalues.  We will consider gauge
group SU(3) and staggered fermions which are related to the chiral
unitary ensemble of RMT.  At zero chemical potential $\mu$, all Dirac
eigenvalues are purely imaginary, and it has been shown in lattice
simulations that $P(s)$ agrees with the Wigner surmise of RMT,
\begin{equation}
  \label{eq1}
  P_{\rm W}(s)=\frac{32}{\pi^2}\,s^2\,e^{-\frac{4}{\pi}s^2}\:,
\end{equation}
both in the confinement and in the deconfinement phase \cite{Hala95}.
This result implies that the Dirac eigenvalues are strongly
correlated, and is indicative of quantum chaos, according to the
conjecture by Bohigas, Giannoni, and Schmit \cite{Bohi84b}.  In
contrast, quantum systems whose classical analogs are integrable obey
a Poisson distribution, $P_{\rm P}(s)=e^{-s}$.

Here, we focus on nonzero chemical potential.  For $\mu\ne0$,
the lattice Dirac matrix generalizes to
\begin{eqnarray*}
  \label{eq2}
  M_{x,y}(U,\mu)\!\!\!&\!=\!&\!\!\!
  \frac{1}{2a} \sum\limits_{\nu=\hat{x},\hat{y},\hat{z}}
  \left[U_{\nu}(x)\eta_{\nu}(x)\delta_{y,x\!+\!\nu}-{\rm h.c.}\right]\\
  &&\!\!\!\!\!\hspace*{-59pt}
  +\frac{1}{2a}\left[U_{\hat{t}}(x)\eta_{\hat{t}}(x)e^{\mu}
    \delta_{y,x\!+\!\hat{t}}
    -U_{\hat{t}}^{\dagger}(y)\eta_{\hat{t}}(y)
    e^{-\mu}\delta_{y,x\!-\!\hat{t}}\right]\:,
\end{eqnarray*}
with the Kawamoto-Smit phases $\eta$.  The eigenvalues of this matrix
are complex.  In this case, $P(s)$ represents the spacing distribution
of nearest neighbors in the complex plane. For each eigenvalue
$z_0$ one has to find the eigenvalue $z_1$ for which $s=|z_1-z_0|$ is
smallest, with a subsequent average over $z_0$.  This definition
assumes that the spectral density is constant over a bounded region in
the complex plane (and zero outside).  Generally, this is not the case
so that an unfolding procedure must be applied, see Sec.~\ref{sec2}.
However, the spectral density of the so-called Ginibre ensemble of RMT,
where real and imaginary parts of the eigenvalues have the same
average size, is constant inside a circle and zero outside,
respectively \cite{Gini65}.  In this case, $P(s)$ is given by
\cite{Grob88}
\begin{equation}
  \label{eq3}
  P_{\rm G}(s)=c \, p(cs)\:,
\end{equation}
with
\begin{displaymath}
  \label{eq4}
  p(s)=2s\lim_{N\to\infty}\left[\prod_{n=1}^{N-1}e_n(s^2)\,e^{-s^2}\right]
  \sum_{n=1}^{N-1}\frac{s^{2n}}{n!e_n(s^2)}\:,
\end{displaymath}
$e_n(x)=\sum_{m=0}^n x^m/m!$, and $c=\int_0^\infty ds \, s \,
p(s)=1.1429...$\\  In contrast, the Poisson distribution in the complex
plane, representing uncorrelated eigenvalues, becomes
\begin{equation}
  \label{eq5}
  P_{\bar{\rm P}}(s)=\frac{\pi}{2}\,s\,e^{-\frac{\pi}{4}s^2}\:.
\end{equation}
In the following, we will study the Dirac spectrum on the lattice at
various values of $\mu\neq 0$ and compare the resulting $P(s)$ with
Eqs.~(\ref{eq3}) and (\ref{eq5}).

\section{Analysis of complex spectra}
\label{sec2}
\vspace{-2mm}
While the quenched
approximation at $\mu\ne 0$ is unphysical \cite{Step96}, there is
currently no feasible solution to the problem of a complex weight
function.  Therefore, we generated the gauge field configurations at
$\mu=0$ and added the chemical potential to the Dirac matrix
afterwards.  The simulations were done with gauge group SU(3) on a
$6^3\times4$ lattice using $\beta=5.2$ and $N_f=3$ flavors of
staggered fermions of mass $ma=0.1$.  For each parameter set, we
sampled 50 independent configurations.

The eigenvalue spectrum is shown in Fig.~\ref{fig1} for five different
values of $\mu$.  The size of the real parts of the eigenvalues grows
with $\mu$ as expected.  While the spectrum is confined to a bounded
region in the complex plane, the spectral density is certainly not
constant.  Therefore, the spectrum has to be unfolded.  In one
dimension, unfolding is a local rescaling of the eigenvalue density
such that the density on the unfolded scale is equal to unity.  To the
best of our knowledge, unfolding in the entire complex plane has not
been considered yet (in Ref.~\cite{Fyod97}, unfolding was done in
strips perpendicular to the real axis).  We proceed as follows.  The
spectral density has an average and a fluctuating part,
$\rho(x,y)=\rho_{\rm av}(x,y)+\rho_{\rm fl}(x,y)$.  Unfolding means to
find a map $z'=x'+iy'=u(x,y)+iv(x,y)$ such that $\rho_{\rm
  av}(x',y')\equiv 1$.  Since the probability has to be invariant,
$\rho_{\rm av}(x',y')dx'dy' = dx'dy' = \rho_{\rm av}(x,y)dxdy$, we
find that $\rho_{\rm av}(x,y)$ is the Jacobian of the transformation
from $(x,y)$ to $(x',y')$,
$  \label{eq6}
  \rho_{\rm av}(x,y)=\left|\partial_xu\,\partial_yv-
    \partial_yu\,\partial_xv\right|\:.
$
Choosing $y'=v(x,y)=y$ yields $\rho_{\rm av}(x,y)=|\partial_xu|$ and,
thus, 
$
  \label{eq7}
  x'=u(x,y)=\int_{-\infty}^xdt\rho_{\rm av}(t,y)\:.
$
Essentially, this is a one-dimensional unfolding in strips parallel to
the real axis.  For a fixed bin in $y$, $\rho_{\rm av}(x,y)$ is
determined by fitting $\rho(x,y)$ to a low-order polynomial.  $P(s)$
is then constructed from the constant unfolded density as explained in
Sec.~\ref{sec1}, normalized such that $\int_0^\infty ds\,s\,P(s)=1$.

\begin{figure*}
  \label{fig2}
  \centerline{\epsfig{figure=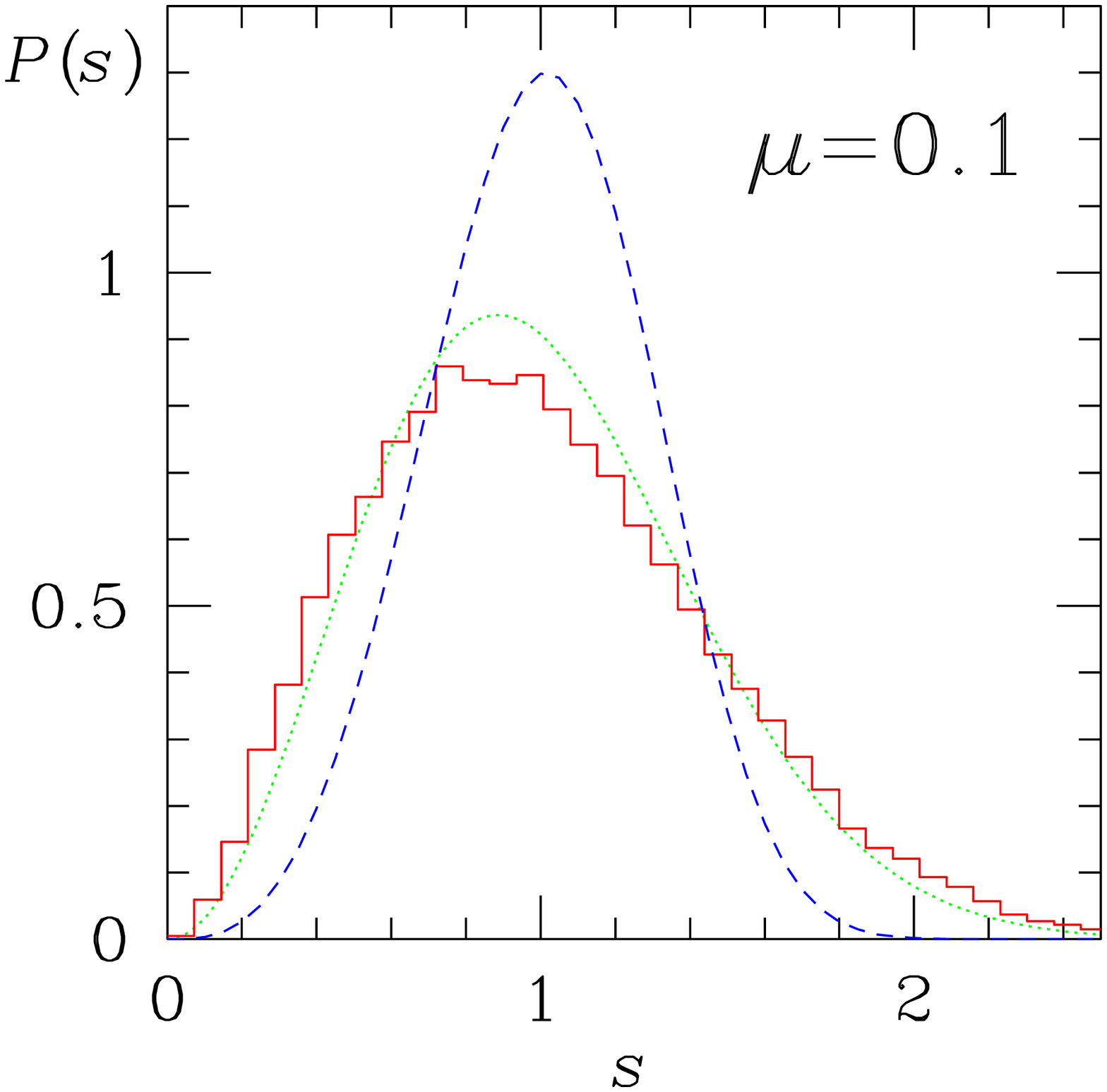,width=30mm}\hspace*{1mm}
    \epsfig{figure=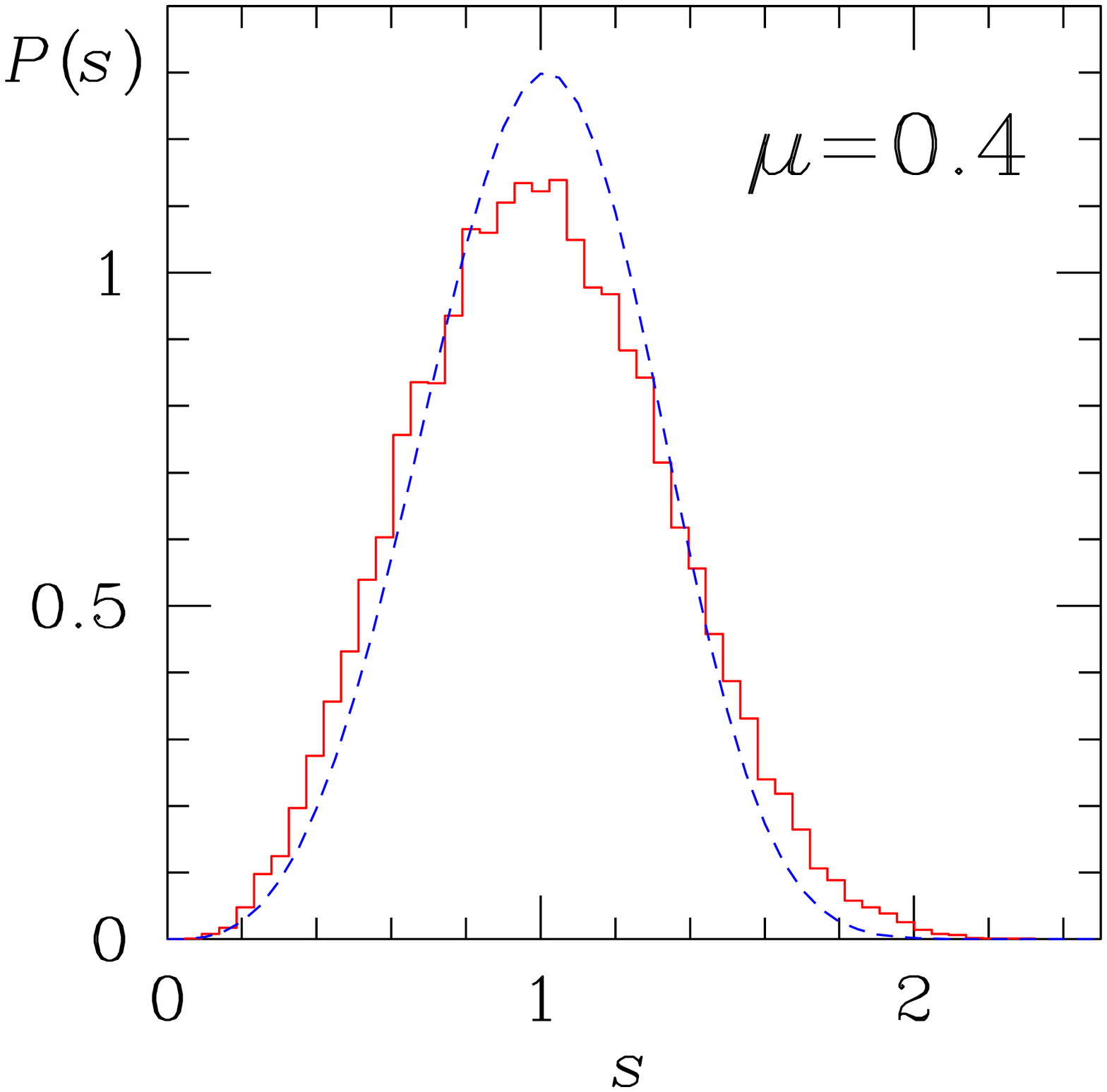,width=30mm}\hspace*{1mm}
    \epsfig{figure=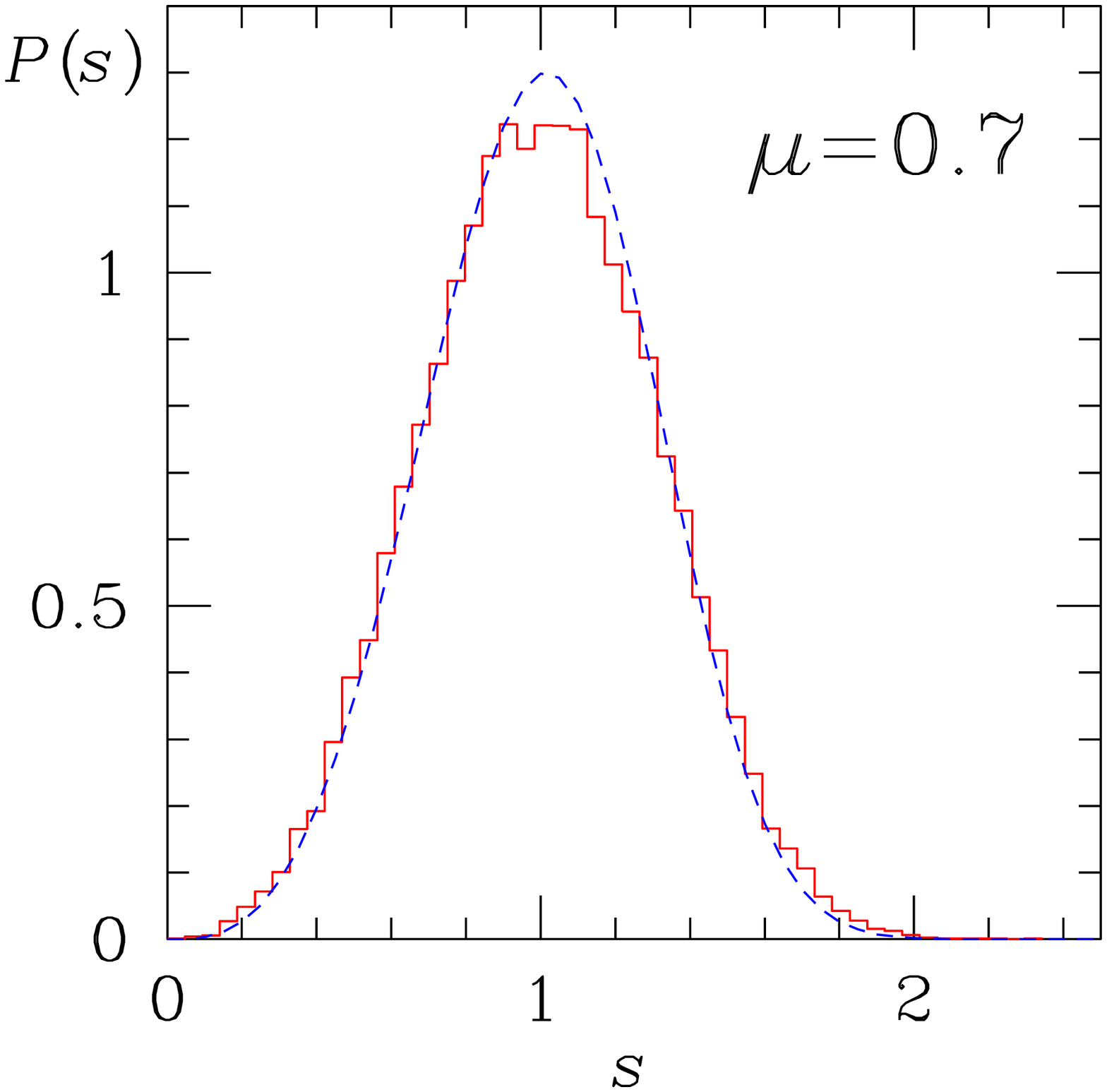,width=30mm}\hspace*{1mm}
    \epsfig{figure=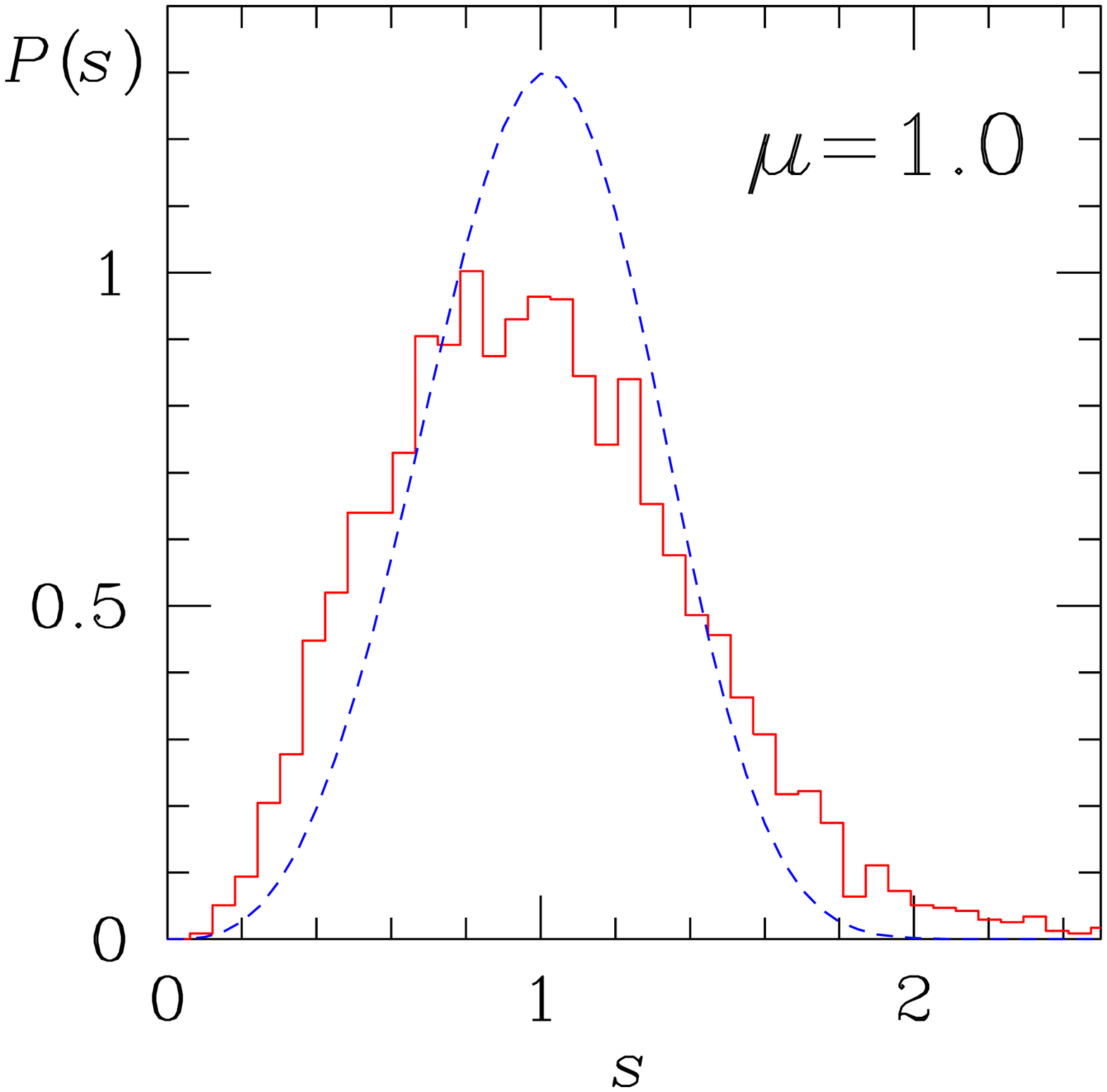,width=30mm}\hspace*{1mm}
    \epsfig{figure=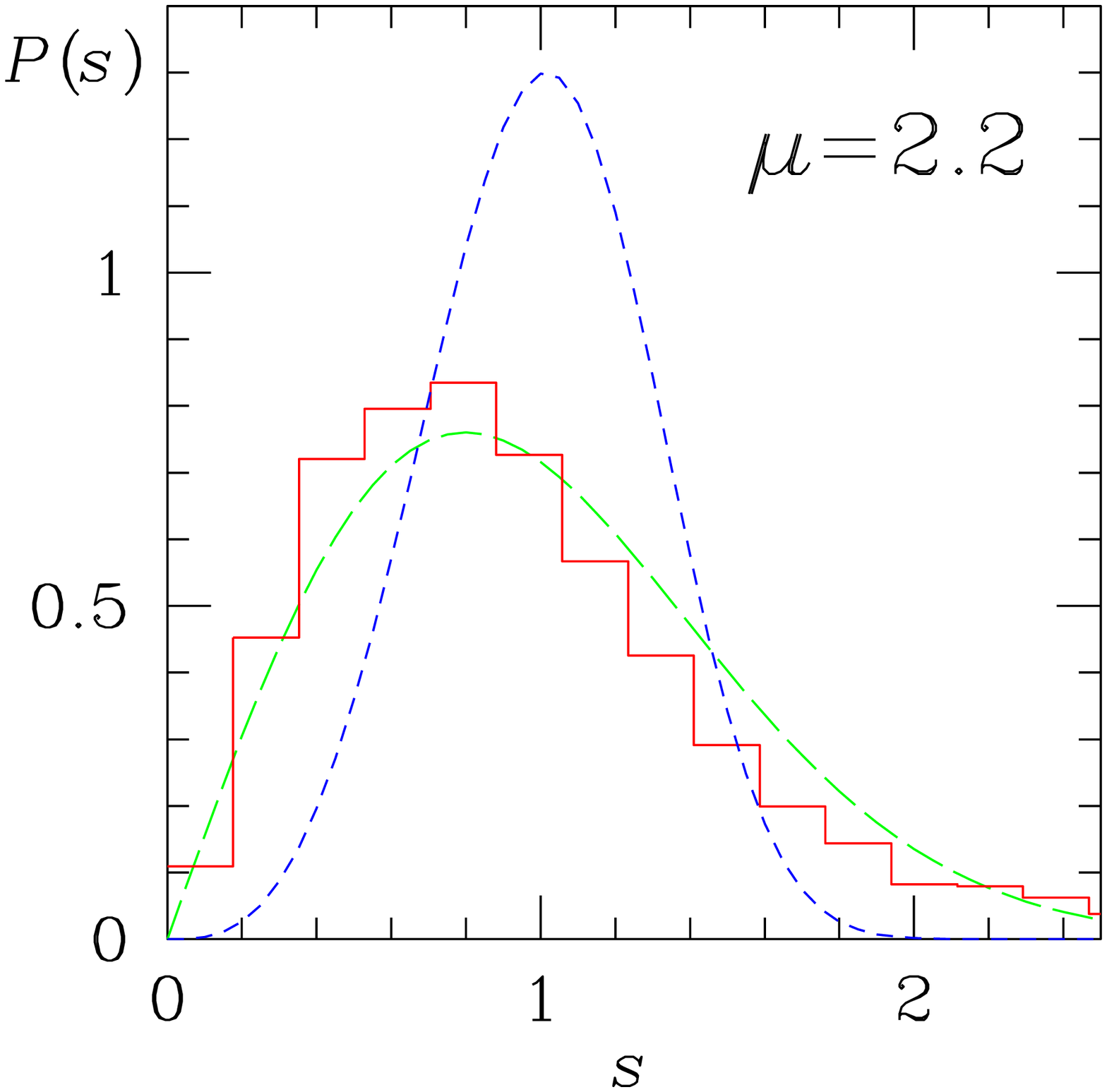,width=30mm}}
  \vspace*{-8mm}
  \caption{Nearest-neighbor spacing distribution of the Dirac operator
    with complex eigenvalues for various values of $\mu$.  The
    histograms represent the lattice data. The short-dashed curve is
    the Ginibre distribution of Eq.~(\protect\ref{eq3}), the dotted
    curve in the left plot the Wigner distribution of
    Eq.~(\protect\ref{eq1}), and the long-dashed curve in the right
    plot the Poisson distribution of Eq.~(\protect\ref{eq5}),
    respectively.}
\end{figure*}

Several remarks are in order.  (i) We have checked spectral
ergodicity.  If only parts of the spectral support are considered, the
results for $P(s)$ do not change.  (ii) If the spectral density has
``holes'' (see Fig.~\ref{fig1} for $\mu=1.0$ and 2.2), we split the
spectral support into several convex pieces and unfold them
separately.  This is justified by spectral ergodicity.  (iii)
Unfolding each spectrum separately and ensemble unfolding yield
the same results for $P(s)$.  (iv) The results for $P(s)$ are stable
under variations of the degree of the fit polynomial and of the bin
sizes in $x$ and $y$.

\section{Results and discussion}
\label{sec3}
\vspace{-2mm}
Our results for $P(s)$ are presented in Fig.~\ref{fig2}.  For small
values of $\mu$, we probe the regime of ``weak non-hermiticity''
\cite{Fyod97} where the typical size of the real part of the
eigenvalues is of the order of the average separation of the imaginary
parts, see Fig.~\ref{fig1} (our definitions differ from
Ref.~\cite{Fyod97} by a factor of $i$ so that it would be more
appropriate to speak of ``weak non-anti-hermiticity'').  Some
analytical results are known \cite{Fyod97} in this regime, but the
spacing distribution is only given as an expansion of $P(s,z_0)$ for
small $s$, where $z_0$ is the location in the complex plane (i.e., no
unfolding is performed).  Therefore, we have no closed analytical
result to compare with, and presumably it would be very difficult to
derive one.  However, something can be said about the small-$s$
behavior of $P(s)$.  For the Ginibre ensemble, we have $P_{\rm
  G}(s)\propto s^3$ for $s\ll1$.  In the regime of weak
non-hermiticity, it is known that $P(s)$ starts with a power smaller
than three \cite{Fyod97}.  This is conform with our results for
$\mu=0.1$ in Fig.~\ref{fig2} where the Wigner surmise $P_{\rm W}(s)$
of Eq.~(\ref{eq1}) for the corresponding hermitian matrix with 
$\mu=0$ is inserted.

Increasing $\mu$, we observe stronger level repulsion leading to quite
nice agreement of the data for $\mu=0.7$ with the prediction from the
Ginibre ensemble.  For this value of $\mu$, we are in the regime of
``strong non-hermiticity'' where the Ginibre ensemble applies.

For values of $\mu$ larger than 0.7, the lattice data deviate
substantially from the Ginibre distribution.  The main question is if
our result for $\mu=2.2$ in Fig.~\ref{fig2} should be interpreted as a
transition to Poisson behavior, or if the agreement of the data at
this particular value of $\mu$ with $P_{\bar{\rm P}}(s)$ of
Eq.~(\ref{eq5}) is purely accidental. We have to keep in mind that we
neglected $\mu$ in the generation of the equilibrium gauge fields.
Therefore, we are not entitled to make any definite statements on
possible connections between the finite density phase transition
(expected at smaller values of $\mu$) and the observed deviations from
Ginibre behavior.  Nonetheless, these deviations are interesting and
deserve further study.

In conclusion, we have investigated the nearest-neighbor spacing
distribution of the lattice Dirac operator with finite chemical
potential.  The data agree with the RMT prediction for the Ginibre
ensemble for $\mu=0.7$ and show deviations for smaller and larger
values of $\mu$.  This is somewhat different from the situation at
$\mu=0$ where the data agree with the RMT results \cite{Hala95}.

\bigskip\noindent{\bf Acknowledgments.} This work was supported in
part by FWF project P10468-PHY and DFG grant We 655/15-1.  We thank
J.J.M. Ver\-baar\-schot for useful discussions.

\end{document}